\documentclass[showpacs,prb,twocolumn,amsmath,floatfix]{revtex4}

\usepackage{graphicx}
\usepackage{dcolumn}

\begin{document}

\newcommand{\alphato}{\alpha_{\rm TO}}
\newcommand{\alphalo}{\alpha_{\rm LO}}
\newcommand{\mutwo}{\mu^{(2)}}
\newcommand{\omegato}{\omega_{\rm TO}}
\newcommand{\omegalo}{\omega_{\rm LO}}
\newcommand{\chitwo}{\chi^{(2)}}
\newcommand{\chiin}{\chi^{(2)}_{\infty}}
\newcommand{\chimw}{\chi^{(2)}_{\rm mw}}
\newcommand{\chieo}{\chi^{(2)}_{\rm eo}}
\newcommand{\pp}{\mathbf{P}}
\newcommand{\ff}{\mathbf{F}}
\newcommand{\uu}{\mathbf{u}}
\newcommand{\ee}{\mathbf{E}}
\newcommand{\eee}{\boldsymbol{\mathcal{E}}}
\newcommand{\p}{\widetilde{p}}

\title{Nonlinear optics of III-V semiconductors in the terahertz
regime: an ab-initio study}

\author{Eric Roman,$^1$ Jonathan R. Yates,$^{1,2}$ Marek Veithen,$^3$
David Vanderbilt,$^4$ and Ivo Souza$^{1,2}$}
\affiliation{$^1$Department of Physics, University of California,
Berkeley, CA 94720\\
$^2$Materials Science Division, Lawrence Berkeley National Laboratory,
Berkeley, CA 94720\\
$^3$D\'{e}partement de Physique, Universit\'{e} de Li\`{e}ge,
B-5, B-4000 Sart-Tilman, Belgium\\
$^4$Department of Physics and Astronomy, Rutgers University,
Piscataway, New Jersey 08854-8019}

\date{\today}

\begin{abstract}

We compute  from first principles the
infrared dispersion of the
nonlinear susceptibility $\chi^{(2)}$ in zincblende semiconductors.
At terahertz frequencies the nonlinear susceptibility depends
not only on the purely electronic response $\chi^{(2)}_{\infty}$,
but also on three other parameters
$C_1$, $C_2$ and $C_3$ describing the contributions from ionic motion.
They relate to the TO Raman polarizability, the second-order
displacement-induced dielectric polarization, and the third-order lattice
potential.
Contrary to previous theory, we find 
that mechanical anharmonicity ($C_3$) dominates over electrical
anharmonicity ($C_2$), which is consistent with recent experiments on GaAs.
We predict that the sharp minimum in the intensity of second-harmonic
generation recently observed for GaAs between
$\omegato/2$ and $\omegato$ does not occur for several other
III-V compounds.

\end{abstract}


\maketitle


\section{INTRODUCTION}
\label{sec:intro}

The nonlinear optical properties of materials in the visible and
near-infrared (IR) have been extensively studied since the development of
the laser.
Except for a few pioneering efforts,\cite{faust66,mayerkeilmann} the far-IR 
region of the
spectrum has remained largely unexplored.
This state of affairs, a consequence of the lack of tunable and intense
laser sources and sensitive detectors in the terahertz range, is starting to
change, thanks to advances in
instrumentation.\cite{davies04,borak05} More accurate measurements
of the nonlinear susceptibilities at terahertz frequencies are beginning to
appear,\cite{dekorsy} calling for quantitative theoretical modeling.

The first nonlinear susceptibility
$\chitwo(\omega_3=\omega_1+\omega_2;\omega_1,\omega_2)$, which is nonzero in
acentric crystals, displays strong dispersion when the
frequencies involved are near a zone-center transverse optical phonon
frequency $\omega_{\rm TO}$. This behavior was
first observed by Faust and Henry\cite{faust66} (see also
Refs.~\onlinecite{faust68} and \onlinecite{barmentlo94}) 
in GaP for the case of mixing between
visible and far-IR radiation ($\omega_1,\omega_1+\omega_2>\omega_{\rm TO}>
\omega_2$); they showed that the dispersion of this process depends on the
linear electro-optic susceptibility $\chieo$.
Another early set of experiments\cite{boyd71,pollack71} investigated
frequency mixing in the microwave range below the lattice resonances
($\omega_1,\omega_2,\omega_1+\omega_2 < \omega_{\rm TO}$).
For the two zincblende compounds studied,
GaAs and GaP, it was found\cite{pollack71}
that the sign of the microwave coefficient
$\chimw$ (the static nonlinear susceptibility) was opposite to
that of the ``high-frequency'' coefficient $\chiin$
describing second-harmonic generation (SHG) and frequency mixing in the
transparency region of the crystal
$(\omega_{\rm TO}<\omega_1,\omega_2,\omega_1+\omega_2< E_g/\hbar)$.
Mayer and
Keilmann\cite{mayerkeilmann} later studied the dispersion of the
SHG coefficient $\chitwo_{\rm SHG}(\omega)=
\chitwo(2\omega;\omega,\omega)$
of GaAs and LiTaO$_3$ over a limited frequency range (0.6 to 1.7~THz).

The present work was motivated by the recent experiments of
Dekorsy {\it et al.}\ on SHG in
GaAs.~\cite{dekorsy}
Using a tunable free-electron laser they
measured the dispersion of $\chitwo_{\rm SHG}$ from 4 to 6~THz, observing a
strong resonant enhancement at 4.5~THz, close to
$\omega_{\rm TO}/2=4.1$~THz as expected, followed by a sharp
dip in the power output. Although they were unable to determine unambiguously
the frequency $\omega_0$ at which $\chitwo_{\rm SHG}$ vanishes, a lower bound of
$5.3$~THz was established. By fitting the location of the minimum in
SHG power to an
expression derived by Flytzanis,\cite{flytzanis72a} they obtained new
parameters for GaAs.

Flytzanis's expression is given below  [Eq.~(\ref{eq:chi_2})]. Using an
effective-bond model, he obtained numerical estimates for its parameters.
These have been used as a guide for interpreting subsequent
experiments,\cite{mayerkeilmann,dekorsy} in spite of their questionable
reliability. For instance, the model predicts the wrong sign for the Born
effective charge and the Raman polarizability, two of the material parameters
at play. In this work we re-examine this problem using first-principles
density-functional techniques.

The paper is organized as follows. In Sec.~\ref{sec:method} we
discuss the formalism describing the IR dispersion of $\chitwo$.
The computational approach is explained in Sec.~\ref{sec:approach}.
In Sec.~\ref{sec:results} we present and discuss results for several III-V
semiconductors,
GaAs, GaP, AlP, AlAs, and AlSb. 
Finally, in Sec.~\ref{sec:summary} we summarize
our findings. Additional discussion of the computational methods is given
in two Appendices.

\section{Formalism}
\label{sec:method}

In this work we limit ourselves to the zincblende structure adopted by
III-V semiconductors, the simplest crystal structure where a 
non-vanishing $\chitwo$ is allowed by symmetry. In zincblende there is a  
single TO mode, and third-rank tensors such as $\chitwo$ have only one 
independent component $\chitwo_{xyz}$. The dispersion of $\chitwo$ below the 
electronic resonances and above the elastic resonances of the medium is given 
by the following expression, obtained by Flytzanis:\cite{flytzanis72a}
\begin{equation}
\chi^{(2)}(\omega_1+\omega_2; \omega_1, \omega_2) = \chiin \,
\Lambda(\omega_1,\omega_2),
\label{eq:chi_2}
\end{equation}
where
\begin{eqnarray}
  \Lambda(\omega_1,\omega_2) &=& 1+C_1 \left(
            \frac{1}{D(\omega_1)} +
            \frac{1}{D(\omega_2)} +
            \frac{1}{D(\omega_1+\omega_2)} \right) \nonumber \\
   &&+C_2 \left(
            \frac{1}{D(\omega_1) D(\omega_2)} +
            \frac{1}{D(\omega_2) D(\omega_1+\omega_2)} \right. \nonumber \\
   &&\qquad\left. +\frac{1}{D(\omega_1+\omega_2) D(\omega_1)} \right) \nonumber \\
   &&+C_3 \left(
            \frac{1}{D(\omega_1) D(\omega_2) D(\omega_1+\omega_2) } \right)
\end{eqnarray}
and $D(\omega) = 1 - \omega^2 / \omegato^2 -
i \gamma \omega / \omegato^2$ is the resonance denominator with phonon damping.
This expression can be derived under rather general 
conditions,\cite{kleinman62} and is independent of the details of the 
microscopic forces, the only assumption being that the mode is
weakly damped ($\gamma \ll \omega_{\rm TO}$).

The dispersion depends on three dimensionless coefficients,
\begin{equation}
\label{eq:c1}
C_1=\frac{\alphato}{2v\chiin}
\left(\frac{Z^*}{M \omegato^2} \right)
\end{equation}
(known as the Faust-Henry coefficient),
\begin{equation}
\label{eq:c2}
C_2=\frac{\mu^{(2)}}{2v\chiin} \left(\frac{Z^*}{M \omegato^2} \right)^2,
\end{equation}
and
\begin{equation}
\label{eq:c3}
C_3=-\frac{\phi^{(3)}}{2v\chi_\infty^{(2)}} 
\left(\frac{Z^*}{M \omegato^2} \right)^3.
\end{equation}
Here $v$ is the volume of the primitive cell,
$M$ is the reduced mass, and the remaining quantities are discussed below.
Unlike the second-rank tensor properties,
the signs of the $C_i$'s remain unchanged if we
reverse the definition of the positive [111] direction.
We adopt the convention that it points from the cation
to the closest anion.\cite{miller70,cardona74}

Having obtained the above model-independent expression for the IR dispersion
of $\chitwo$, Flytzanis then proceeded to estimate the values of the 
coefficients $C_1$, $C_2$, and $C_3$ for several III-V compounds,
using an effective-bond model.\cite{flytzanis72a} We will now describe
how to compute them from first-principles.
Except for the damping parameter $\gamma$ (which we do not calculate),
the quantities entering Eqs.~(\ref{eq:chi_2}--\ref{eq:c3}) are conveniently
evaluated as derivatives of forces or macroscopic
polarization with respect to macroscopic electric fields or displacements
(forces under finite fields are readily available via the
Hellmann-Feynman theorem\cite{souza02}).
In what follows $\ff$ represents the force on the cation, $\pp$ is the
macroscopic polarization, $\eee$ is a macroscopic electric field, and
$\uu=\uu_{\rm III}-\uu_{\rm V}$ is
the relative displacement between the cation (group~III) and anion (group~V)
sublattices away from their equilibrium positions.
Two of the quantities in Eqs.~(\ref{eq:chi_2}--\ref{eq:c3}) are first
derivatives: the cation Born effective charge
\begin{equation}
\label{eq:eff_charge}
Z^*=v\left.
  \frac{\partial P_x}{\partial u_x}\right|_{\ee=0}=
  \left.\frac{\partial F_x}{\partial {\cal E}_x}
\right|_{\uu=0},
\end{equation}
and the zone-center TO phonon frequency
\begin{equation}
\label{eq:freq}
\omegato^2=-\frac{1}{M}
\left.\frac{\partial F_x}{\partial u_x}\right|_{\eee=0}.
\end{equation}
The remaining four are second derivatives:
the non-resonant electronic (``high-frequency'') quadratic
susceptibility
\begin{equation}
\label{eq:chiin}
\chiin=
\left.
  \frac{1}{2}\frac{\partial^2 P_x}{\partial {\cal E}_z\partial {\cal E}_y}
\right|_{\uu=0},
\end{equation}
the non-resonant TO Raman polarizability per primitive cell\footnote{The 
quantity $\alphato$ corresponds to
   $\widetilde{\alpha}^{(1)}$ of
   Refs.~\onlinecite{flytzanis71} and \onlinecite{flytzanis72a}, and
   $-{\cal P}$ of Refs.~\onlinecite{cardona74} and
\onlinecite{varshney81}.}
\begin{equation}
\label{eq:raman}
\alphato=
\left.
v\frac{\partial[\chi^{(1)}_\infty]_{xy}}{\partial u_z}\right|_{\eee=0}=
\left.
  \frac{\partial^2 F_z}{\partial {\cal E}_x\partial {\cal E}_y}
\right|_{\uu=0},
\end{equation}
the second-order dipole moment or ``electrical anharmonicity''
\begin{equation}
\label{eq:mu2}
\left.
  \mu^{(2)}=v\frac{\partial^2 P_x}{\partial u_z\partial u_y}
\right|_{\eee=0},
\end{equation}
and the third-order lattice potential or ``mechanical
anharmonicity''
\begin{equation}
\label{eq:phi3}
\phi^{(3)}=
\left.
-\frac{\partial^2 F_x}{\partial u_z\partial u_y}
\right|_{\eee=0}.
\end{equation}
Note that in Eq.~(\ref{eq:raman}) the Raman
tensor was recast
as a second-order field-induced force.\cite{komornicki79,jackson00}

It will be useful to consider
Eq.~(\ref{eq:chi_2}) in the three limiting cases discussed in the 
Introduction. In the high-frequency limit it 
reduces to the purely electronic coefficient $\chiin$;
for $\omega_1,\omega_1+\omega_2 > \omegato > \omega_2$ it becomes the
unclamped-ion, strain-free electro-optic coefficient, 
\begin{equation}
\label{eq:chieo}
\chieo = \chiin ( 1+C_1 );
\end{equation}
and finally for $\omega_1, \omega_2,\omega_1+\omega_2 < \omegato $ it
describes the strain-free static (microwave) nonlinear susceptibility, which 
involves all three coefficients,
\begin{equation}
\label{eq:chimw}
\chimw = \chiin ( 1+3C_1+3C_2+C_3 ).
\end{equation}

\section{Computational approach}
\label{sec:approach}

The calculations are done using
{\tt ABINIT}, \cite{gonze02} 
a plane-wave pseudopotential
density-functional code, using both the local-density approximation (LDA)
and a generalized gradient approximation\cite{perdew1} (GGA). 
In order to better assess the
sensitivity of the nonlinear optical properties to the approximate density
functional used, all calculations
are performed at the same (experimental) lattice constants.
Norm-conserving Troullier-Martins pseudopotentials\cite{troullier1} are used
for all materials, and
for Ga the $d$-electrons are included in the valence.
We used a cutoff energy of 30~Ha for the aluminum compounds, and 45~Ha for the
gallium compounds.

For finite systems such as molecules, {\it ab-initio} calculations of
$\chitwo$ below the electronic resonances (including contributions from ionic
motion\cite{kirtman97}) are performed routinely.
Similar calculations for bulk solids have become feasible only recently, thanks
to a series of developments, starting with the Berry-phase theory of
polarization,\cite{king-smith93} which provides a means of evaluating the
electronic contribution to the macroscopic polarization $\pp$ as a bulk
quantity
(for a review, see Ref.~\onlinecite{vanderbilt06}).
In Ref.~\onlinecite{veithen05} a density-functional perturbation
method was developed for computing nonlinear (electronic,
electro-optic, and nonresonant Raman) susceptibilities using the
Berry-phase formalism
to treat the electric field perturbation. Here we use a closely-related
approach where the derivatives are evaluated by finite differences
rather than analytically, using the method of Ref.~\onlinecite{souza02} to
handle finite electric fields.
The appeal of this method lies in its
simplicity: once implemented, no extra coding
is required to compute a different higher-order or
mixed derivative, or to switch from, e.g., LDA to GGA.

In our calculations
we apply the finite perturbation ($\uu$ or $\eee$) along [111], and monitor
the response ($\pp$ or $\ff$) along the same direction.  That is, we let
$\uu=\delta a(\hat{\bf x}+\hat{\bf y}+\hat{\bf z})$,
where $a$ is the lattice constant and the cation and anion sublattices are
brought closer together when $\delta>0$;
$\eee=\varepsilon E_0(\hat{\bf x}+\hat{\bf y}+\hat{\bf z})$,
where $E_0 = e/(4 \pi \epsilon_0 a_0^2)$;
$\ff=f(\hat{\bf x}+\hat{\bf y}+\hat{\bf z})$; and
$\Delta\pp=p(\hat{\bf x}+\hat{\bf y}+\hat{\bf z})$.
Eqs.~(\ref{eq:eff_charge}--\ref{eq:phi3}) then become
\begin{equation}
Z^*=
\frac{a^2}{4}\frac{\partial p}{\partial\delta}=
\frac{\partial f}{\partial\varepsilon},
\end{equation}
\begin{equation}
\omegato^2=
-\frac{1}{aM}\frac{\partial f}{\partial\delta},
\end{equation}
\begin{equation}
\chiin=
\frac{1}{4}\frac{\partial^2 p}{\partial\varepsilon^2},
\end{equation}
\begin{equation}
\alphato=
\frac{1}{2}\frac{\partial^2 f}{\partial \varepsilon^2},
\end{equation}
\begin{equation}
\mu^{(2)}=\frac{a}{8}\frac{\partial^2 p}{\partial \delta^2},
\end{equation}
and
\begin{equation}
\phi^{(3)}=
-\frac{1}{2a^2}\frac{\partial^2 f}{\partial \delta^2}.
\end{equation}
The parameters are calculated from these expressions, and
then inserted into Eqs.~(\ref{eq:c1}--\ref{eq:c3}) to obtain the coefficients
$C_1$, $C_2$, and $C_3$.
In Appendix~\ref{sec:app_direct} we describe a different approach whereby 
$\chieo$ and $\chimw$  are evaluated directly by finite differences
in addition to $\chiin$, and Eqs.~(\ref{eq:chieo}) and
(\ref{eq:chimw}) are then used to obtain $C_1$ and $3C_2+C_3$.

We have taken as the smallest increments $\delta=1\times10^{-3}$ and 
$\varepsilon=1\times10^{-4}$ for AlAs, AlP, GaP, and AlSb.
In the case of GaAs a smaller field step of
$\varepsilon=3\times10^{-5}$
was used. This was needed in order to stay below the $k$-mesh
dependent critical field above which the electric enthalpy functional
loses its minima.\cite{souza02} Because of its smaller band gap, the
critical field is lower for GaAs than for the other compounds.

For the derivatives we use Richardson's extrapolation to estimate the
limit $h\to0$ from calculations with two different step sizes:
\begin{equation}
\label{eq:richardson}
f^{(n)}(x) = \frac{4}{3} D^{(n)}(x, h) - \frac{1}{3} D^{(n)}(x, 2h)+
{\cal O}(h^4),
\end{equation}
where $D^{(1)}$ is given by
the centered finite difference expression
\begin{equation}
D^{(1)}(x,h) \equiv \frac{f(x+h) - f(x-h)}{2h}=f'(x)+{\cal O}(h^2),
\end{equation}
and $D^{(2)}$ is given by
\begin{eqnarray}
D^{(2)}(x,h)&\equiv& \frac{f(x+h) + f(x-h) - 2f(x)}{h^2}\nonumber\\
&=&f''(h)+{\cal O}(h^2).
\end{eqnarray}
In order to speed up the convergence of polarization-dependent quantities
with respect to the $k$-point sampling, we use a similar extrapolation
for the discretized Berry-phase formula, as described in
Appendix~\ref{sec:app_extrap}. All the values quoted in the tables were
calculated on a $16\times 16\times 16$ $k$-point mesh, except in the
case of GaAs which, for reasons discussed in that Appendix, demanded a denser 
mesh.

\section{Results for III-V Semiconductors}
\label{sec:results}

\subsection{Microscopic parameters}

\begin{table}
\caption{Parameters that determine the nonlinear susceptibility of zinclende 
compounds in the infrared range, grouped into first- and
second-derivative quantities 
[Eqs.~(\ref{eq:eff_charge}--\ref{eq:phi3})]. Rows labelled ``Model''
pertain to the empirical-model calculations of Flytzanis:
$\chiin$ is taken from Ref.~\onlinecite{flytzanis68}, the remaining
values from Ref.~\onlinecite{flytzanis72a}.}
\label{table:matsecondorder}
\begin{ruledtabular}
\begin{tabular}{ll|dd|D{.}{}{1}D{,}{}{1}dd}
& & \multicolumn{2}{c|}{First derivative} & \multicolumn{4}{c}{Second derivative}\\
 & &
 \multicolumn{1}{c}{$Z^\star$} &
 \multicolumn{1}{c|}{$\omega_{\rm TO}$} &
 \multicolumn{1}{c}{$\chiin$} &
 \multicolumn{1}{c}{$\alphato$} &
 \multicolumn{1}{c}{$\mu^{(2)}$} &
 \multicolumn{1}{c}{$\phi^{(3)}$}\\
 &
 &
 &
 \multicolumn{1}{c|}{(THz)} &
 \multicolumn{1}{c}{$\mathrm{(pm/V)}$} &
 \multicolumn{1}{c}{$\mathrm{(\AA^2)}$} &
 \multicolumn{1}{c}{$\mathrm{(nC/m)}$} &
 \multicolumn{1}{c}{$\mathrm{(TJ/m^3)}$}\\
\hline
GaAs &LDA      &2.05  &8.0 &472   &-54    &-1.89&4.29\\
     &GGA      &2.05  &8.3 &337   &-34    &-1.67&5.10\\
     &Model  & &  &127&84&5.90   &-1.2\\
     &Expt.     &2.2\footnotemark[1]   &8.0\footnotemark[1]   &170.\footnotemark[2] &-38,\footnotemark[3]& &\\
\hline
GaP  &LDA      &2.10  &10.6 &131 &-12,.9 &-1.50   &4.82\\
     &GGA      &2.15  &10.9 &114 &-9,.8  &-1.36   &4.76\\
     &Model  &-2.0&11.5      & 93 &43&7.77   &-2.1\\
     &Expt.      &2.0\footnotemark[1]   &11.0\footnotemark[1] & 71.\footnotemark[2] &-20,\footnotemark[3]& &\\
\hline
AlP  &LDA      & 2.24 &12.7&45   &-5  &0.28    &4.64\\
     &GGA      &2.24  &13.1&42   &-5  &0.24    &4.32\\
     &Expt.    &2.28\footnotemark[1]  & 13.2\footnotemark[4]& & & &\\
\hline
AlAs &LDA       & 2.14 &10.4 &79  &-9  &0.27    &4.00\\
     &GGA      &2.11  &10.8 &73  &-8  &0.22    &4.94\\
     &Expt.     &2.3\footnotemark[1]   & 10.8\footnotemark[5]    &    &      &        &\\
\hline
AlSb &LDA      &1.86  &9.2 &205 &-19 &-0.09   &3.09\\
     &GGA      &1.79  &9.5 &187 &-18 &-0.12   &3.34\\
     &Model    &-1.6&9.8 & 47&77& 8.34    & -1.5\\
     &Expt.    &1.9\footnotemark[1]   &9.6\footnotemark[1] & 153.\footnotemark[2]   &      &        &\\
\end{tabular}
\end{ruledtabular}
\footnotetext[1]{Quoted in Ref.~\onlinecite{yuandcardona03}.}
\footnotetext[2]{Ref.~\onlinecite{landoltbornstein}.}
\footnotetext[3]{Ref.~\onlinecite{varshney81}.}
\footnotetext[4]{Quoted in Ref.~\onlinecite{rodriguez89}.}
\footnotetext[5]{Quoted in Ref.~\onlinecite{gianozzi91}.}
\end{table}

We have systematically computed
the values of all six parameters (\ref{eq:eff_charge}--\ref{eq:phi3}), 
using the methods summarized above,
for five III-V compounds. The results are collected in 
Table~\ref{table:matsecondorder}, and we will begin by discussing the
comparison with the model theory of Flytzanis, and then pass to the
discussion of the experimental values.

Table~\ref{table:matsecondorder} shows striking differences between the
first-principles results and Flytzanis's model calculations: (i) While both 
levels of theory produce the same sign for $\chiin$, they disagree on the 
signs of $Z^*$, $\alphato$, $\mu^{(2)}$, and $\phi^{(3)}$. (ii) We find that 
the magnitude of $\mu^{(2)}$ ($\phi^{(3)}$) is significantly smaller (larger) 
than Flytzanis predicts.

Born charges and phonon frequencies are routinely computed from 
first-principles, and they tend to compare favorably with 
experiment,\cite{gianozzi91} as evidenced in
Table~\ref{table:matsecondorder}. Dielectric susceptibilities and
Raman polarizabilities are more problematic. For example, it is well-known
that density-functional theory tends to overestimate the dielectric constant.
This also seems to be generally the case for the nonlinear susceptibility 
$\chiin$.\cite{levineallan,dalcorso} 
The problem here is compounded by the
fact that the experimental determination
of this quantity is also problematic, with the values
reported in the literature displaying a very large 
dispersion.\cite{varshney81} We have opted for quoting the recommended values 
from Landolt-B\"ornstein.\cite{landoltbornstein} 
Inspection of Table~\ref{table:matsecondorder} suggests that
our first-principles values are too large by roughly a factor of two,
which however is comparable with the uncertainty in the experimental 
determination. Measurements of the absolute Raman polarizability are also
difficult, and few values are reported in the literature. 
Our result that $\alphato<0$ means
that the bond polarizability along [111] increases with increasing
bond length around the equilibrium length of the bond. This is in
agreement with the measured sign in GaAs,\cite{cardona74} but in 
disagreement with the model
calculations of Flytzanis.\cite{flytzanis71,flytzanis72a} 

\subsection{Lattice-induced contributions to $\chitwo$}

>From the calculated parameters in Table~\ref{table:matsecondorder} we obtained,
using Eqs.~(\ref{eq:c1}--\ref{eq:c3}), the various lattice-induced
contributions to $\chitwo$, collected in Table~\ref{table:c1c2c3}. 
We find that 
the Faust-Henry coefficients $C_1$ are roughly a factor of two
smaller than the experimental values, 
consistent with the overstimation of $\chiin$ in Eq.~(\ref{eq:c1})
discussed above.
The mechanical anharmonicity coefficient $C_3$ is,
in all cases, significantly larger in magnitude than $C_2$ (electrical
anharmonicity). This is the opposite conclusion from
Refs.~\onlinecite{flytzanis72a} and \onlinecite{flytzanis72b}.
We remark that although the correct signs for $C_1$ and $C_3$ were obtained
therein, this resulted
from a cancellation of errors in the signs of $Z^*$, $\alpha_{\rm TO}$, and
$\phi^{(3)}$ in Eqs.~(\ref{eq:c1}) and (\ref{eq:c3}). No such cancellation
occurs in $C_2$, and indeed for the three compounds studied both in the present
work and in Ref.~\onlinecite{flytzanis72a} (AlSb, GaP and GaAs), there is a
disagreement in the predicted sign for this quantity.

While
the coeffficient $C_1$ can be measured in various
ways (electro-optic effect, frequency mixing,\cite{faust66} and relative
Raman scattering efficiencies from LO and TO
phonons\cite{johnston69,cardona82}) with fairly consistent results,
it is difficult to disentangle the values of $C_2$ and $C_3$ from
experiments. The more readily accessible quantity is the combination
$3C_2+C_3$: it follows from Eqs.~(\ref{eq:chieo}) and (\ref{eq:chimw}) that
\begin{equation}
\label{eq:3c2+c3}
3C_2 + C_3 = 2 + \frac{\chimw - 3 \chieo}{\chiin},
\end{equation}
where all the quantities on the right-hand-side are directly measurable.
Flytzanis found $C_2>0$ and $C_3<0$ for all the III-V compounds he
investigated. Under those circumstances,
the measured sign of $3C_2 + C_3$ indicates which anharmonic
contribution (electrical or mechanical)
is dominant in a given material. We find, however, that the sign of
$C_2$ is not the same
for all III-V compounds, which invalidates such an analysis.

\begin{table}
\caption{Lattice-induced contributions to the nonlinear susceptibility of 
zincblende compounds [Eqs.~(\ref{eq:c1}--\ref{eq:c3})].
Rows labelled ``Model''
pertain to the empirical-model calculations of Flytzanis.
}
\label{table:c1c2c3}
\begin{ruledtabular}
\begin{tabular}{llddd}
     &       &    \multicolumn{1}{c}{$C_1$}  &   \multicolumn{1}{c}{$C_2$}    
     &   \multicolumn{1}{c}{$C_3$}    \\
\hline
   AlP &LDA &-0.38 &0.05 & -1.82\\
       &GGA &-0.37 &0.04 &-1.78\\
\hline
  AlAs &LDA &-0.37 &0.03 &-0.91\\
       &GGA &-0.34 &0.02 &-0.82\\
\hline
  AlSb & LDA & -0.25 & -0.00 & -0.22 \\
       & GGA & -0.23 & -0.00 &-0.18 \\
       &Model&-1.97 &0.35 &-0.11\\
\hline
   GaP &LDA &-0.27 &-0.04 &-0.53\\
       & GGA & -0.28 & -0.07 & -0.56\\
       & Model & -0.37 & 0.11 & -0.05\\
       & Expt. &-0.53\footnotemark[1]&\\
       & Expt. &-0.75\footnotemark[1]& &\\
\hline
  GaAs &LDA &-0.35 &-0.02 &-0.12\\
       &GGA &-0.29 &-0.03 &-0.15\\
       &Model&-0.83 &0.14 &-0.07\\
       &Expt. &-0.51\footnotemark[1] & &\\
       &Expt. &-0.59\footnotemark[2] & &\\
       &Expt. &-0.68\footnotemark[1] & &\\
       &Expt. &-0.48\footnotemark[3] & &\\
\end{tabular}
\end{ruledtabular}
\footnotetext[1]{Quoted in Ref.~\onlinecite{flytzanis72a}.}
\footnotetext[2]{Ref.~\onlinecite{johnston69}.}
\footnotetext[3]{Ref.~\onlinecite{martini71}.}
\end{table}

Interestingly, our calculated
$3C_2+C_3$ disagree in sign with the values inferred from
experiment\cite{boyd71,pollack71}
(see Table~\ref{table:sum_and_ratio}).
It is apparent from Eq.~(\ref{eq:3c2+c3}) that the sign
of $3C_2+C_3$ is rather sensitive to not only the signs, but also the relative
magnitudes of $\chieo$ and $\chimw$. While the signs of our calculated
$\chimw$ and $\chieo$ agree with experiment
(see Table~\ref{table:sum_and_ratio})
there is a significant discrepancy regarding their magnitudes.
The well-known limitations of density-functional theory in reproducing
dielectric properties, such as the optical gap (underestimated) and the
the dielectric constant (overestimated)
may be of concern in this
regard. We note however that a possible error in the magnitude of
$\chiin$ will not affect the sign of $3C_2+C_3$, while $\phi^{(3)}$, $Z^*$,
and $\mu^{(2)}$, the remaning parameters entering $C_2$ and $C_3$, are expected
to be reasonably accurate within density functional theory, which typically
describe rather well
lattice-dynamical and zero-field polarization properties (e.g., Born
charges) of III-V semiconductors.\cite{gianozzi91}
Hence our
prediction for the sign for $3C_2+C_3$ should be sound.
In view of the discrepancy with experiment, it would be useful to
have careful measurements of the relative magnitudes of
$\chimw $ and $\chieo$, but we are not aware of any other work along these
lines besides the pioneering investigations of
Boyd {\it et al.}\cite{boyd71,pollack71}

A convenient measure of the relative importance of the two 
lattice-anharmonicity mechanisms is the ratio $C_2/C_3$, also included in
Table~\ref{table:sum_and_ratio}. We expect
reasonably accurate {\it ab-initio} results for this quantity, since it is
independent of $\chiin$, the ``weak link''
in the calculation.
Our values clearly cannot be reconciled with those of Flytzanis.
In the next section we will discuss what this implies for
the interpretation of the recent experiment of Dekorsy
{\it et al.},\cite{dekorsy} which
attempted to obtain values for the parameters $C_2$ and $C_3$
separately for the first time.

\begin{table}
\caption{Third and fourth columns:
parameters $C_2$ and $C_3$ combined in a way that relates more
directly to experiments. Fifth and sixth columns: electro-optic and
microwave nonlinear suceptibilities of Eqs.~(\ref{eq:chieo})
and (\ref{eq:chimw}) (in pm/V). Rows labelled ``Model''
pertain to the empirical-model calculations of Flytzanis.}
\label{table:sum_and_ratio}
\begin{ruledtabular}
\begin{tabular}{llddD{.}{}{1}D{.}{}{1}}
 &   &      \multicolumn{1}{c}{$3C_2+C_3$}  &  
\multicolumn{1}{c}{$C_2/C_3$}&    \multicolumn{1}{c}{$\chieo$}  &  
\multicolumn{1}{c}{$\chimw$}\\
\hline
AlP  &LDA&    -1.68  & -0.026& 28 & -82 \\
     &GGA&    -1.66  & -0.022&       26   & -74\\
\hline
AlAs &LDA&    -0.83  & -0.028&       50   & -75\\
     &GGA&    -0.77  & -0.023&       48   & -58\\
\hline
AlSb &LDA&    -0.23  &  0.012&       154  &   6\\
     &GGA&    -0.19  &  0.017&       145  &  25\\
     &Model& 0.93 &-3.33\\
\hline
GaP  &LDA&    -0.71  &  0.146&       88   & -90\\
     &GGA&    -0.78  &  0.128&        82  & -69\\
     &Model& 0.27 &-2.22\\
     &Expt.&    0.28\footnotemark[1]  &&   20.\footnotemark[1]  & -24.\footnotemark[1]\\
\hline
GaAs &LDA&    -0.19  &  0.203&    309  & -107\\
     &GGA&    -0.22  &  0.175&    240  &-30\\
     &Model& 0.35 &-1.96\\
     &Expt.&    0.39\footnotemark[1]  &&  43.\footnotemark[1]  & -51.\footnotemark[1]\\
\end{tabular}
\end{ruledtabular}
\footnotetext[1]{Refs.~\onlinecite{boyd71,pollack71}.}
\end{table}

\subsection{Zero-crossings of $\chitwo_{\rm SHG}$ in the terahertz range}

\begin{figure}
\includegraphics[width=\columnwidth]{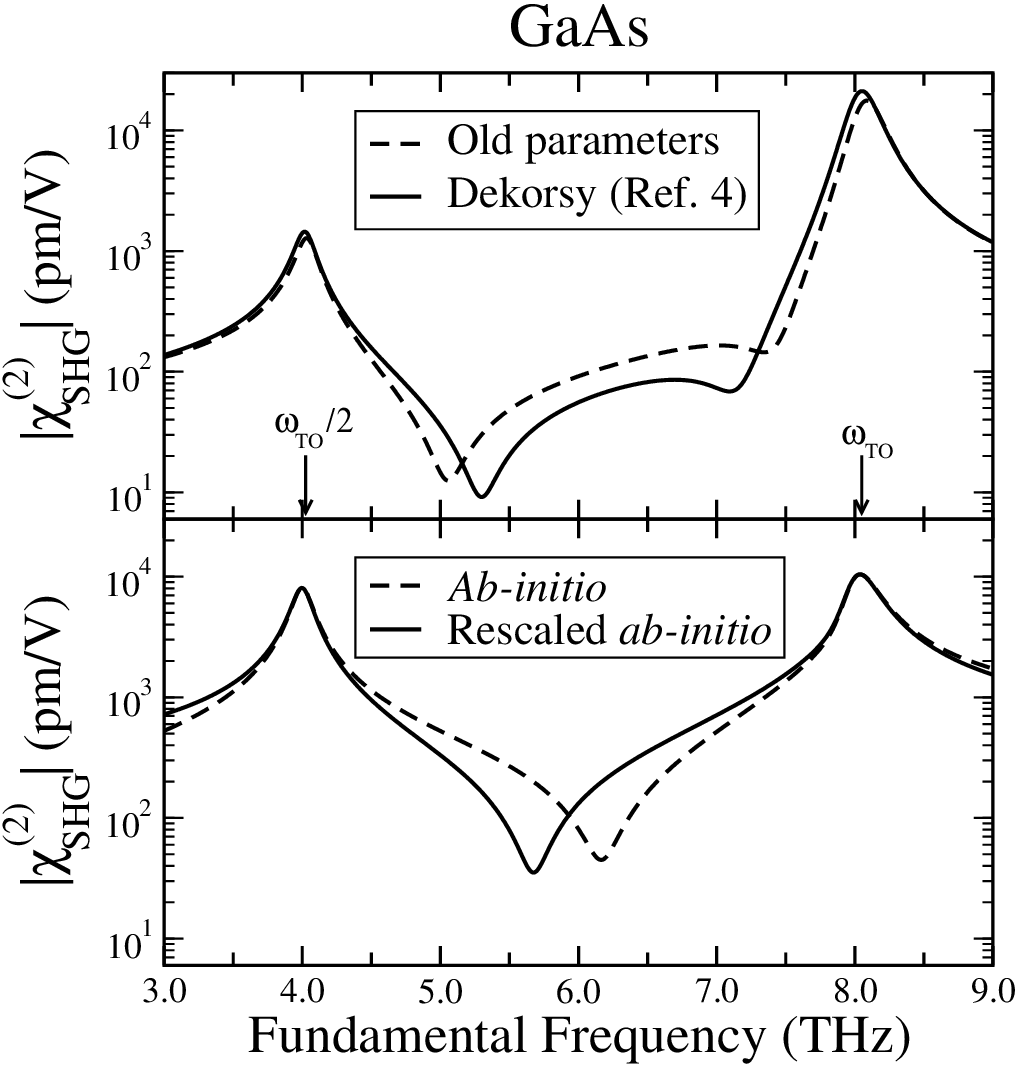}
\caption{
Calculated IR dispersion of $|\chitwo_{\rm SHG}(\omega)|$ in GaAs, 
for different
choices of the parameters in Eq.~(\ref{eq:chi_2}). 
$(C_1,C_2,C_3)=(-0.59,0.14,-0.07)$, $(-0.59,0.16,-0.13)$, 
$(-0.35,-0.024,-0.12)$
and $(-0.59,-0.041,-0.20)$ for the curves labeled `Old parameters,' 
`Dekorsky,'
`{\it Ab-initio},' and `Rescaled {\it ab-initio},' respectively.  
Following Ref.~\onlinecite{dekorsy}, we set the
damping parameter $\gamma$ to 0.29~THz. The meaning
of these parameter sets is explained in the text.
}
\label{fig:chi2_gaas}
\end{figure}

The quantity $|\chitwo_{\rm SHG}(\omega)|$ is displayed in
Fig.~\ref{fig:chi2_gaas} for GaAs.
The dashed line in the upper panel correspond to
a sensible choice of parameters assembled from the experimental and
theoretical investigations from the 1960's and 1970's
(these will be 
referred to as ``old parameters'').
In between the expected strong resonant enhancements at $\omegato/2$ and
$\omegato$, there are two dips, at 5.1 and 7.4~THz, the first
more pronounced than the second. They result from
sign reversals of ${\rm Re}\,\chitwo_{\rm SHG}(\omega)$ in
Eq.~(\ref{eq:chi_2}). If SHG is observed over
a frequency range containing the first zero crossing, its frequency
$\omega_0$ can be detected as a sharp dip in the second harmonic power.
If, furthermore, $3C_2+C_3$ is known from separate measurements of
$\chiin$, $\chieo$ and $\chieo$,\cite{boyd71,pollack71}
the remaining free parameter in Eq.~(\ref{eq:chi_2}),
$C_2/C_3$, can then be adjusted to fit the zero-crossing frequency.
This was proposed in Ref.~\onlinecite{mayerkeilmann} as a way of
determining $C_2$ and $C_3$ separately.

Dekorsy {\it et al.}\cite{dekorsy} recently used a free-electron laser to
measure the far-IR dispersion of the SHG power in GaAs from 4.4 to 5.6~THz.
They observed the expected resonance close to $\omegato/2$, followed by a
strong drop. Because of insufficient filtering of the fundamental signal in the
detector above 5.6~THz, which masked the SHG signal, they were unable to
locate precisely the zero crossing, and only a
{\it lower bound} of 5.3~THz was established. Since this is slightly above
the 5.1~THz predicted for $\omega_0$ from the old parameters, they then
discussed how the values of $C_2$ and $C_3$ had to be revised to increase
$\omega_0$ to 5.3~THz (the assumption being that
the lower bound is reasonably close to the actual zero-crossing).
They opted to leave $3C_2+C_3$ unchanged at 0.35 (the
theoretical value from Ref.~\onlinecite{flytzanis72a}, which is fairly close
to the experimental 0.39); a good fit, shown as a
solid line in the upper panel of Fig.~\ref{fig:chi2_gaas},
was then obtained by
changing $C_2/C_3$ from $-2.0$ to about $-1.23$. This amounts to essentially
doubling $C_3$, from $-0.07$ to $-0.14$, while changing $C_2$ only slightly,
from 0.14 to 0.16.

\begin{figure}
\includegraphics[width=\columnwidth]{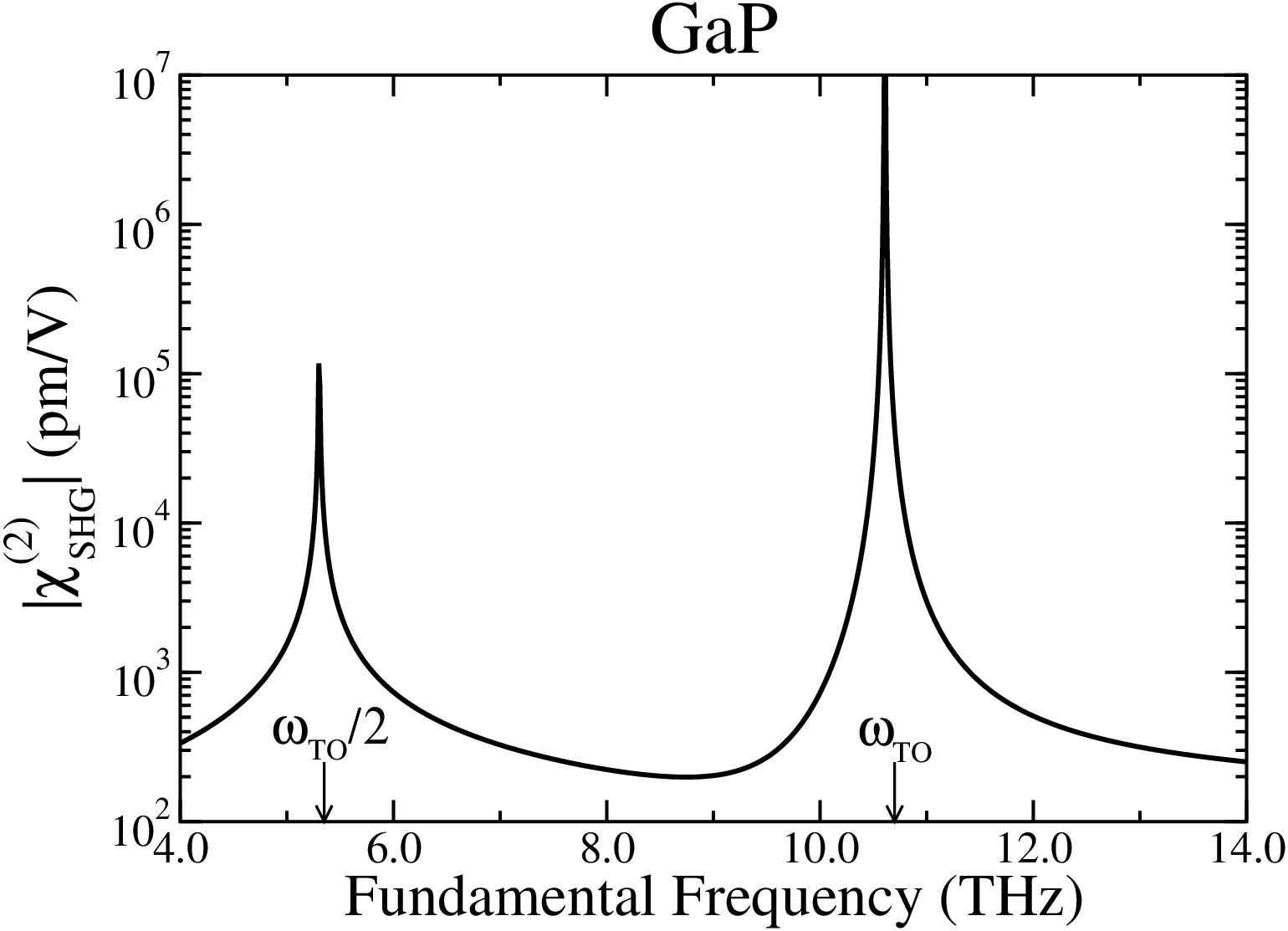}
\caption{
Calculated IR dispersion of $|\chitwo_{\rm SHG}|$ in GaP, using the
{\it ab-initio} parameters without damping ($\gamma = 0~\mathrm{THz}$).
In contrast to GaAs (Fig.~\ref{fig:chi2_gaas}),
no sharp dip is observed between the two maxima at
$\omegato/2$ and $\omegato$.}
\label{fig:chi2_gap}
\end{figure}

The dashed line in the lower panel of Fig.~\ref{fig:chi2_gaas}
shows the dispersion obtained
with our {\it ab-initio} parameters. The zero-crossing frequency is raised
significantly, to 6.2~THz.
In order to assess the impact of the
uncertainty in $\chiin$ on the dispersion, we show as a solid line the curve 
that results from reducing $\chiin$ from 472~pm/V to 277~pm/V.
This affects the $C_i$'s according to 
Eqs.~(\ref{eq:c1}--\ref{eq:c3}), and we have chosen the amount of rescaling 
so as to bring our value for $C_1$ into agreement with the
experimental number from Ref.~\onlinecite{johnston69}, $C_1=-0.59$.
The zero-crossing frequency also changes, from 6.2~THz to 5.67~THz,
only slightly above the measured lower bound of 5.3~THz.
Clearly, different sets of values for $C_2$ and
$C_3$  can lead to dispersions with very similar zero crossings
(the solid lines in the two panels of Fig.~\ref{fig:chi2_gaas}), and thus
both consistent with the
experimental data, underscoring the need for reliable theoretical input.
Interestingly, we find that for the other III-V compounds no zero
crossing occurs for $\omegato/2 < \omega < \omegato$. 
This is illustrated in Fig.~\ref{fig:chi2_gap} for GaP.

\section{Summary}
\label{sec:summary}

We have carried out a detailed {\it ab-initio} investigation of
the IR dispersion of the non-linear susceptibility $\chitwo$ in III-V
zincblende semiconductors. The results were compared with Flytzanis's
empirical model\cite{flytzanis72a} and with experiment,
with particular emphasis on the recent second-harmonic generation measurements
carried out by Dekorsy {\it et al.}\cite{dekorsy} These authors based the 
interpretation of their data on the parameters obtained in 
Ref.~\onlinecite{flytzanis72a} from model calculations. By revising them 
somewhat, they were able to obtain a reasonable fit to the IR dispersion of
$\chitwo_{\rm SHG}$. Instead, we find a completely different set of 
parameters, which however is still consistent with experiment.

Our findings can be summarized as follows: (i) We provide theoretical support
to the main qualitative conclusion of Ref.~\onlinecite{dekorsy}, that the ratio
$|C_2/C_3|$ between
the contribution from  second-order lattice dipole moment ($C_2$) versus
phonon interaction through the third-order lattice potential
anharmonicity ($C_3$) is
smaller than previously thought. (ii) However, we find that this is a 
consequence of not
only an increase in $|C_3|$,\cite{dekorsy} but also a significant
decrease in $|C_2|$, with the result that the former dominates the latter
($|C_2/C_3|\ll 1$). (iii) The sign of $C_2$ is not constant thoroughout the
III-V series, and for the two most-studied compounds (GaAs and GaP) it is
negative, contrary to prior understanding.
(iv) For all compounds except AlSb, we find that the sign of the microwave
nonlinear susceptibility $\chimw$
is opposite to that of the optical ($\chiin$) and electro-optical ($\chieo$)
coefficients, in agreement with early experiments on GaAs and
GaP.\cite{boyd71,pollack71}  However, our calculated negative
sign for $3C_2+C_3$, which is sensitive to the relative
magnitudes of $\chieo$ and $\chimw$, disagrees with those experiments.
(v) The parameter fit to the IR dispersion of $\chitwo_{\rm SHG}$
relies on the occurrence of a sharp
minimum in SHG power between $\omegato/2$ and $\omegato$.
The observed zero-crossing in GaAs is reproduced by our calculations, but is
also consistent with an alternative set of parameters characterized by
$3C_2+C_3>0$ and $|C_2/C_3|>1$.
For the other compounds considered, we find no zero-crossing.

\section{Acknowledgments}

This work was supported by the Laboratory Directed Research and 
Development Program of Lawrence
Berkeley National Laboratory under the Department of Energy Contract
No. DE-AC02-05CH11231 and by NSF Grant DMR-0549198.

\appendix

\section{A method for the direct evaluation of $C_1$ and $3C_2+C_3$}
\label{sec:app_direct}

As discussed in the main text, by using Eqs.~(\ref{eq:chieo}) and 
(\ref{eq:chimw})  one can obtain
experimental values for
$C_1$ and $3C_2+C_3$ from separate measurements of $\chiin$, $\chieo$, and
$\chimw$.
In this Appendix we describe the computer-experiment analogs of such
measurements.

We define ``high-frequency'' (optical)
fields $\eee$ and ``low-frequency'' (static) fields $\ee$
as follows:\cite{veithen05}
when an optical field $\eee$ is applied the ions are not allowed
to move, so that the polarization response is purely electronic; in contrast,
both ions and electrons respond to a static field $\ee$, resulting in a total
change in polarization that contains ionic as well as electronic contributions
(but the cell is kept
strain-free, since we are interested in frequencies above the elastic
resonances of the medium). In Ref.~\onlinecite{souza02} the
high-frequency and static dielectric constants, $\epsilon_\infty$
and $\epsilon_0$, were evaluated by finite differences as first
derivatives with respect to $\eee$ and $\ee$, respectively.
Extending this to second order yields the
high-frequency $\chiin$ and the low frequency  (microwave) $\chimw$.

Evaluating the electro-optic $\chieo$ requires
combining a static and an optical field:
\begin{equation}
\label{eq:chi2_eo}
\big[\chieo\big]_{xyz}=\frac{1}{2}\frac{d}{dE_z}
\left(\frac{\partial P_x}{\partial{\cal E}_y}\right)=
\frac{1}{2}\frac{d}{dE_z}
\big[\chi_\infty^{(1)}\big]_{xy}.
\end{equation}
Here the action of the static field is described by a total derivative as a
reminder that the
polarization depends on a static field both explicitly and implicitly, through
the atomic positions. Clearly the order of the (partial and total) derivatives
matters. We can view their combined action as a conventional
mixed derivative on an
auxiliary function $\widetilde{\pp}(\ee,\eee)$ defined as follows:
(i) apply a field $\ee$ and let both electrons and
ions respond; (ii) add a  field $\eee$, let the
electrons readjust under the total field $\ee+\eee$ while keeping the ions
fixed in the positions obtained
in the first step. $\widetilde{\pp}(\ee,\eee)$ is defined as the polarization
after step (ii). Then
\begin{equation}
\frac{d}{dE_z}
\left.\Big(\frac{\partial P_x}{\partial{\cal E}_y}\Big)\right|_{\ee=\eee=0}=
\frac{\partial^2\widetilde{P}_x}{\partial E_z\partial{\cal E}_y}=
\frac{\partial^2\widetilde{P}_x}{\partial {\cal E}_y\partial E_z}.
\end{equation}
As before, we apply small fields along [111]:
$\ee=\epsilon(\hat{\bf x}+\hat{\bf y}+\hat{\bf z})$, and
$\eee=\varepsilon(\hat{\bf x}+\hat{\bf y}+\hat{\bf z})$.
Then, defining
$\Delta\widetilde{\pp}(\ee,\eee)=
\widetilde{p}(\epsilon,\varepsilon)(\hat{\bf x}+\hat{\bf y}+\hat{\bf z})$,
we find
\begin{equation}
\chieo=\frac{1}{4}\frac{\partial^2\widetilde{p}}
{\partial\epsilon\partial\varepsilon},
\end{equation}
which we evaluate as
\begin{eqnarray}
\frac{\partial^2\widetilde{p}}
{\partial\epsilon\partial\varepsilon}&=&
\frac{\p(\epsilon,\varepsilon)-\p(\epsilon,-\varepsilon)-
\p(-\epsilon,\varepsilon)+\p(-\epsilon,-\varepsilon)}{4\epsilon\varepsilon}
\nonumber\\
&+&
{\cal O}(\epsilon^2,\varepsilon^2).
\end{eqnarray}
We found that a more stringent force tolerance for the atomic relaxations
must be used when evaluating $\chimw$
than when evaluating $\chieo$. 
This results from the fact that in the latter
case the displacements are second-order in the
field, whereas in the former they are first-order.\cite{rabe02}
Well-converged values were obtained by using
a force tolerance of $10^{-7}$~Ha/bohr for $\chimw$, 
while for $\chieo$ $10^{-5}$~Ha/bohr was sufficient.

The values for $\chieo$ and $\chimw$ obtained using this method
agree to within
$1~\mathrm{pm/V}$ with the ones obtained from the separate calculation of $C_1$,
$C_2$ and $C_3$
described in Sec.~\ref{sec:approach}. This provides an internal
consistency 
check of 
our calculations. Although it is somewhat more expensive and does not
provide as much information (e.g., it does not produce separate
values for $C_2$ and $C_3$), the method described in this Appendix
provides a simple means of
controlling the
mechanical boundary conditions under applied fields. Although we only
considered atomic displacements, the same strategy
can be extended to strain deformations. In this way one can easily
compute, for example,
the clamped (strain-free) and unclamped (stress-free)
electro-optic coefficients, or the static $\chitwo$ including the strain
response.\cite{rabe02}

\section{Improving the convergence with respect to $k$-point sampling}
\label{sec:app_extrap}

Although total-energy ground state calculations for insulators converge 
exponentially fast with respect to $k$-point sampling, for 
finite-field calculations the convergence is considerably
slower.\cite{umari03} This results
from the discretized Berry-phase (DBP) polarization expression
(Eq.~(\ref{eq:secondorderp}) below) used
in
the electric enthalpy functional,\cite{souza02} and
as a consequence
the second field-derivatives in Eqs.~(\ref{eq:chiin}) and (\ref{eq:raman}) 
also converge slowly. In order
to alleviate this problem we used in our finite-field calculations a modified 
DBP expression for the polarization, Eq.~(\ref{eq:fourthorderp}) below.

For notational simplicity we limit our discussion to the case of a single 
valence band in one dimension.
The electronic polarization of a bulk insulator under periodic boundary 
conditions can be written, by analogy with the dipole moment of a molecule,
as $P_{\rm el} = -e \langle x \rangle /L$, where $L$ is the length of the
periodic box.
The Berry-phase expression for $\langle x\rangle$ is\cite{resta98}
\begin{equation}
\label{eq:secondorderx}
\langle x \rangle \simeq
    \frac{L}{2\pi}
    \mathrm{Im}~\ln\langle\Psi|e^{i\frac{2\pi x}{L}}|\Psi\rangle,
\end{equation}
where $\Psi$ is a many-body insulating wavefunction.
For a finite system in a supercell,
evaluating the dipole moment with this expression amounts to 
replacing the non-periodic operator $x$ with the periodic operator
$(L/2\pi) \sin(2\pi x/L)$.\cite{stengel06}
The difference between the two near the origin, where the molecule is located,
is of order $1/L^2$.
Using for $\Psi$ a Slater determinant of single-particle Bloch states,
we recover from this expression the King-Smith-Vanderbilt DBP
expression for the polarization of a band insulator:\cite{resta98}
\begin{equation}
\label{eq:secondorderp}
P_{\rm el} = - \frac{e}{2\pi}
    \mathrm{Im~ln~} \prod_{s=0}^{N-1} \mathrm{det~} S(k_s, k_{s+1})
    + {\cal O}(1/L^2),
\end{equation}
where $S(k_s, k_{s+1})=\langle u_{k_s}|u_{k_s+1}\rangle$.
An alternative expression for $\langle x\rangle$ is 
\begin{equation}
\label{eq:fourthorderx}
\langle x \rangle \simeq
    \frac{L}{2\pi} \left [
    \frac{4}{3}\mathrm{Im}~\ln\langle\Psi|e^{i\frac{2\pi x}{L}}|\Psi\rangle -
    \frac{1}{6}\mathrm{Im}~\ln\langle\Psi|e^{i\frac{4\pi x}{L}}|\Psi\rangle
    \right ]
    ,
\end{equation}
which is correct to ${\cal O}(1/L^4)$, as can be seen using the same type of
reasoning as in Ref.~\onlinecite{stengel06}.
Eq.~(\ref{eq:fourthorderx}) leads to a modified DBP polarization formula,
\begin{eqnarray}
\label{eq:fourthorderp}
P_{\rm el} = - \frac{e}{2\pi}
    &\Big[&
    \frac{4}{3}
    \mathrm{Im~ln~} \prod_{s=0}^{N-1} \mathrm{det~} S(k_s, k_{s+1})
    -\nonumber\\
    &&\frac{1}{6}
    \mathrm{Im~ln~} \prod_{s=0}^{N-1} \mathrm{det~} S(k_s, k_{s+2})
    \Big]
    + {\cal O}(1/L^4)
    .\nonumber\\
\end{eqnarray}
This is closely related to the expression obtained by combining Richardson's
extrapolation, Eq.~(\ref{eq:richardson}), with
Eq.~(\ref{eq:secondorderp}), viewed as a finite-difference representation of
the $k$-derivative in the continuum Berry-phase 
formula.\cite{bennetto96}

\begin{figure}
\vspace{0.25cm}
\includegraphics[width=\columnwidth]{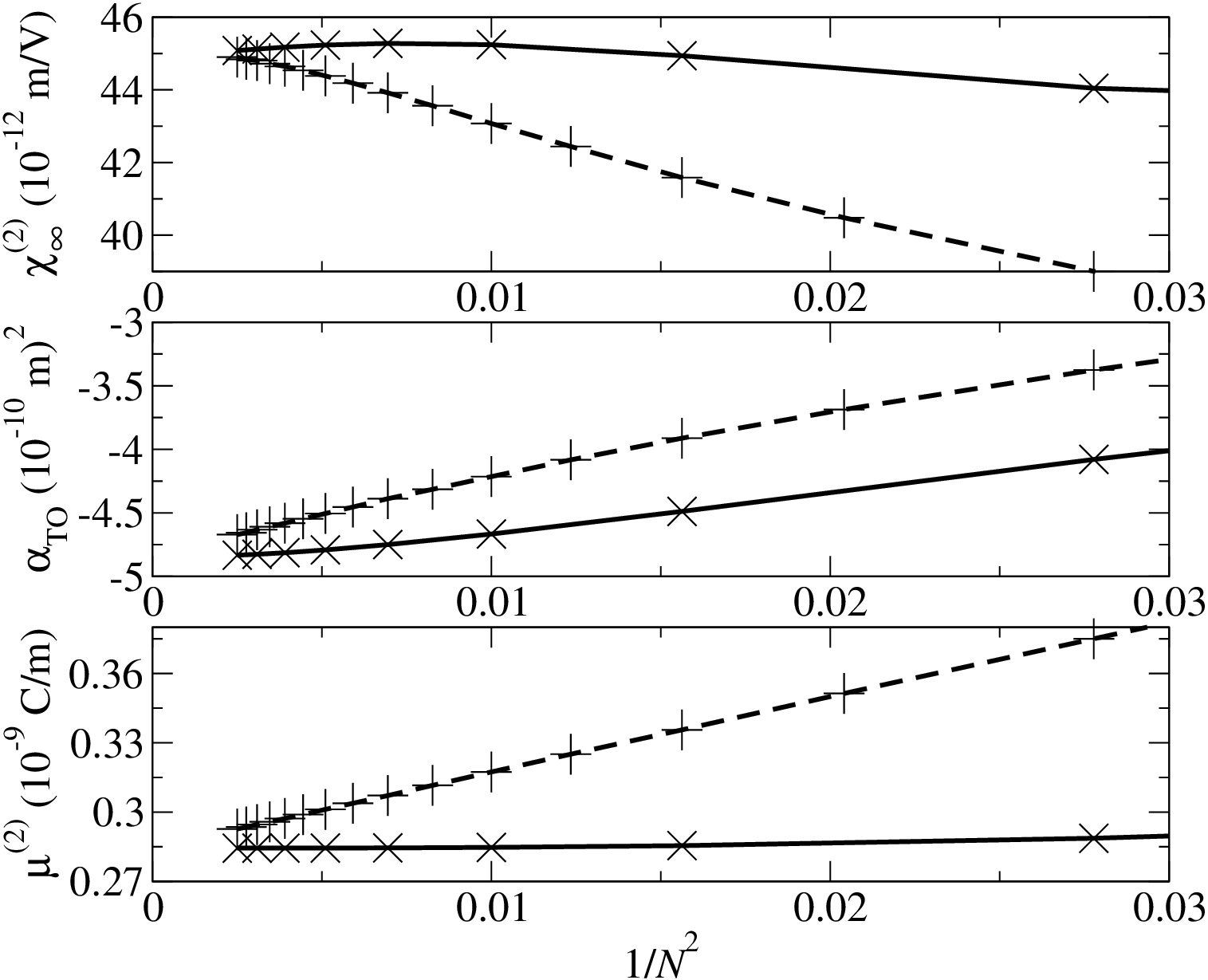}
\caption{Comparison of the
convergence of $\chiin$, $\alphato$, and $\mutwo$ for AlP with respect to
the $k$-point mesh when using the conventional discretized Berry-phase 
expression (\ref{eq:secondorderp}) (dashed lines) versus using the modified 
expression (\ref{eq:fourthorderp}) (solid lines).
}
\label{fig:secondorder}
\end{figure}

We show in Figure~\ref{fig:secondorder}
the convergence with the size $N\times N\times N$ of the shifted Monkhorst-Pack
grid of the second-order quantities $\chiin$, $\alphato$, and $\mutwo$
(these are the third derivatives of the electric enthalpy which involve at 
least one field derivative).  
For each quantity, we plot as a function of $1/N^2$  
the value obtained using both the conventional polarization expression 
(\ref{eq:secondorderp}) (dashed line)
and the modified expression (\ref{eq:fourthorderp}) (solid line).
Notice that the values for $\chiin$, $\alphato$, and 
$\mutwo$ obtained from the former fall
nearly on a straight line.  Extrapolating to $N=\infty$
through a least squares fit against $1/N^2$ is a reliable way of predicting
the converged values, with errors of usually around 1\%.  This procedure
requires several calculations at different $N$, starting at $N=12$ to avoid
the contribution from higher-order terms in $1/N$.  The modified
polarization expression produces results of similar accuracy with a calculation
for a single value of $N$.
We found that $N=16$ usually provides accurate values for the calculation
of nonlinear susceptibilities to within $1~\mathrm{pm/V}$.
As we can see from the graph, $\mutwo$ calculated in this way
converges rapidly:  The resulting value at $N=6$ is closer to the converged 
value of
$0.284~\mathrm{nC/m}$ than that calculated with the conventional
functional at $N=20$.  The improvement for $\alphato$ and $\chiin$ is less
dramatic, but in general there is a clear improvement 
for most materials.  The notable exception is GaAs, the material with the
smallest gap, where there is an 
improvement for all quantities except $\chiin$, which actually
converges more slowly
when using the modified fourth-order expression.
We have found empirically that this expression 
works best for large band gap materials.

\bibliographystyle{apsrev}
\bibliography{infrared,electronicstructure}

\end{document}